\documentclass{JHEP3}
\usepackage{epsfig}
\usepackage{graphicx}
\usepackage{dcolumn}
\usepackage{bm}
\newcommand{\be}{\begin{equation}}
\newcommand{\ee}{\end{equation}}
\def\bea{\begin{eqnarray}}
\def\eea{\end{eqnarray}}

 \def\be{\begin{equation}}
\def\ee{\end{equation}}
\def\bea{\begin{eqnarray}}
\def\eea{\end{eqnarray}}

\def\lesssim{\mathrel{\hbox{\rlap{\hbox{\lower4pt\hbox{$\sim$}}}\hbox{$<$}}}}
\def\gtrsim{\mathrel{\hbox{\rlap{\hbox{\lower4pt\hbox{$\sim$}}}\hbox{$>$}}}}

\title{ New
 Attractors}

\author{ Renata Kallosh
\\
    Department of Physics, Stanford University, Stanford, CA 94305,
USA and \\
Kyoto University, Yukawa Institute, Kyoto, 606-8502, Japan}
\received{\today}       
 \preprint{YITP-05-48 }

\abstract{We derive new algebraic attractor equations describing supersymmetric flux vacua of  type IIB string theory. The first term in these equations, proportional to the gravitino mass (the central charge), is similar to the attractor equations for moduli fixed by the charges near the horizon of the supersymmetric black holes. The second term does not have a counterpart in the theory of black hole attractors. It is proportional to a mass matrix  mixing axino-dilatino with complex structure modulino. This allows stabilization of moduli for vanishing central charge, which was not possible for BPS black holes. Finally, we propose a new set of attractor equations for non-supersymmetric black holes and for non-supersymmetric flux vacua.
}

\begin{document}


\section{Introduction}
    
The supersymmetric attractor equations \cite{Ferrara:1995ih}-\cite{Ferrara:1996um} have been discovered in the context of the theory of BPS black holes. It was found there that the scalar fields near the black hole horizon have a fixed point behavior defined by the black hole electric and magnetic changes. The BPS black holes have been  extensively discussed in string theory context in \cite{Moore:2004fg}, \cite{Denef:2001xn}. 
Non-supersymmetric black hole attractors were introduced  in \cite{Ferrara:1997tw}. They have been rediscovered and studied in many interesting examples more recently in  \cite{Goldstein:2005hq},\cite{Tripathy:2005qp}. The authors have clarified  special features of such non-supersymmetric attractors both in supergravity and in string theory.  In this note we will provide a new form of the black hole attractor equations and show that there is a direct relation between the mathematical structure established for  black hole attractors and for flux vacua. On the other hand, we will show that the attractor equations for the flux vacua has some new parts which are not present in the black hole case. Finally, we will derive  simple attractor equations for non-supersymmetric black holes and flux vacua.

A significant progress in investigation of attractors in string theory was achieved by Moore in \cite{Moore:2004fg}.\footnote{For an early attempt to relate the minimum of the superpotential to the $SL(2,Z)$ black hole attractors see \cite{Kallosh:1996xa}; for a more recent discussion on flux vacua and black hole attractors see \cite{Curio:2000sc}.}
However, his investigation revealed a difficulty which did not allow an immediate generalization of the results for black hole attractors to the study of flux vacua. He  pointed out that in type IIB string theory compactified on a CY$_3$ manifold the flux vacua are described by the complex non-integral $\tau$-moduli-dependent 3-form flux $G_3= F_3-\tau H_3$. The appearance of the field $\tau$ in $G_{3}$  is the source of a problem when one tries to use the black hole attractor equations based on a set of real magnetic and electric charges.
It was  suggested in \cite{Moore:2004fg} that this issue is easier to resolve in the context of M-theory compactified on a CY$_4=$CY$_3\times T^2$ manifold. Moore has developed these ideas for a particular example of $K3\times T^2$. 

Recently there was a renewal of  the interest in black hole attractors. A conjecture was proposed in  \cite{Ooguri:2004zv} about the relation between black hole attractors and topological strings. This conjecture  involves a development in black hole attractors with  higher order corrections. Such corrections were  derived in \cite{LopesCardoso:1998wt} where also the generalized black hole attractor mechanism was suggested. 

In this paper we will establish new attractor equations for the general flux vacua of type IIB string theory compactified on CY orientifolds. We choose O3/O7 orientifolds to be specific, as in the GPK-KKLT vacua \cite{Giddings:2001yu}, \cite{Kachru:2003aw}.

The essential  ingredient in deriving new flux vacua attractor equations which we will use in this paper is a Hodge-decomposition of the space of allowed 4-form fluxes proposed in \cite{Denef:2004ze}.\footnote{I am grateful to F. Denef for pointing out ref. \cite{Denef:2004ze} to me. } The complete decomposition will give an additional term in the attractor equation which is absent for BPS black holes.

We will show that: 
\begin{enumerate}
  \item For generic CY$_3$ manifold with O3/O7 planes one can rewrite the standard {\it differential} equations for supersymmetric flux vacua in IIB string theory, $DW=0$ with  $W\neq 0$, as  {\it algebraic} attractor equations relating fluxes to fixed moduli,  including all complex structure moduli as well as the axion-dilaton. Equation $DW=0$ determines a position of a supersymmetric extremal point of the N=1 supergravity potential,  $V= e^K(|D W|^2-3 |W|^2)$. 
The attractor equations for supersymmetric flux vacua resemble 
the black hole attractor equations \cite{Ferrara:1995ih}-\cite{Ferrara:1996um} with the central charge $Z$ identified with $e^{K\over 2} W$.  For these equations to be valid in flux vacua one has to require in addition  that the second derivative of the central charge (the mass of the chiral fermions)  vanishes in the supersymmetric vacua, $M_{AB}= D_{A}D_{B} Z=0$. For the BPS  black hole case in N=2 supergravity this condition is always satisfied,  see for example eq. (55) in \cite{Ferrara:1997tw}. This is in accordance with the rules of special geometry \cite{deWit:1984pk}, \cite{Ceresole:1995ca}.

  \item  In a more general case when the mass of the chiral fermions does not vanish in the supersymmetric vacua,  $M_{AB}\neq 0$, which signals the deviation from the special geometry,  we will identify the generalized attractor equations, defining the fixed moduli in N=1 flux vacua of type IIB string theory compactified on a CY$_3$ orientifold. The new attractor equations will acquire, in addition to the usual term proportional to the central charge (gravitino mass),  a new term proportional to the masses of chiral fermions.
The new feature of supersymmetric flux vacua, as compared to black hole attractors, is the fact that for $DW=0$  one can fix the moduli  even for $Z=e^{K\over 2} W=0$,  
see e.g. \cite{Giryavets:2003vd}. This is different from the black hole case, when for $Z=0$ one typically finds a null singularity in the geometry and some of the moduli blow up.

\item There are simple attractor equations for non-supersymmetric black holes, which are defined by the minimum of the ``black hole potential'' \cite{Ferrara:1997tw},   \cite{Goldstein:2005hq},  
\be V_{BH}= |DZ|^2+|Z|^2 \ .
\ee 
We also find the corresponding attractor equations for the non-supersymmetric flux vacua defined by the extremal point of the potential
\be V=e^K(|D W|^2-3 |W|^2)=  |D Z|^2-3 |Z|^2 \ .
\label{N1pot}\ee

\end{enumerate}

The (complex) gravitino mass  and the (complex) mass matrix element of chiral fermions in an arbitrary point of the moduli space $(t, \bar t)$ in  string theory compactified into effective 4d N=1 supergravity are 
\begin{equation}
  M_{3/2}= e^{K\over 2} W =\int_X G_4\wedge \Omega_4 \equiv  Z \, [t , \bar t ; p, q]\ ,
\label{gravitino}
\end{equation}
\begin{equation}
  M_{AB} = D_A D_B M_{3/2}= D_{A}D_B \, Z\,  [t , \bar t ; p, q]\ .
\label{mix}
\end{equation}
Here $X$ is   an elliptically fibered CY 4-fold ${ T^2\times Y\over Z_2}$, which allows us to study IIB string theory compactified on CY orientifold.
Here we remind that in effective N=1 supergravity the gravitino mass is defined via the superpotential and the K\"ahler potential as in (\ref{gravitino}), whereas the chiral fermion mass matrix is  $M_{AB}= D_A D_B (e^{K\over 2} W)$.
In string theory compactified on CY with fluxes, the fermion masses depend not only on moduli  but also on fluxes, which we denote by $p,q$. At the supersymmetric attractor point, where
\begin{equation}
  D Z= \bar D \bar Z=0\ ,
\label{susy}
\end{equation}
moduli are fixed by fluxes. They become certain function of fluxes, $t_{fix}(p,q) , \bar t_{fix}(p,q)$. Therefore the fermion masses at the supersymmetric critical point depend only on fluxes:
\begin{equation}
  M_{3/2}^{cr}(p,q)= Z [t(p,q) , \bar t(p,q) ; p, q] \ ,
\label{fixgravitino}
\end{equation}
\begin{equation}
  M_{AB}^{cr}(p,q) = D_{A}D_B Z [t(p,q) , \bar t(p,q) ; p, q] \ .
\label{fixmix}
\end{equation}

An unusual situation for flux vacua with regard to black hole attractors  is that we are looking for N=1 flux vacua  with the N=1 effective potential 
$
   V= e^K(|D W|^2-3 |W|^2) $.
The supersymmetric vacua of our interest are defined by equations $DW=0$,  so the potential at the critical point with $\partial  V=0$ is given by
$
   V_{cr}= -3 e^K|W|_{cr}^2
$.
In cases when $|W|_{cr}$ vanishes, we find Minkowski vacua, otherwise we have AdS vacua.
Our goal here is to show that instead of solving {differential} equations on the superpotential, $DW=0$, the corresponding critical point of the potential can be found   by solving {algebraic} attractor equations relating fluxes to fixed moduli. Note that this is by no means a property of a general N=1 supergravity. An important ingredient of our construction  follows from compactified type IIB string theory on CY$_3$ orientifold: on CY$_3$ manifold we  get N=2 supergravity upon compactification.  An orientifolding, which truncates it to N=1,  does not change the N=2 properties of the geometry of the moduli space surviving orientifolding. After orientifolding, some fields are truncated from the theory. However, for the remaining moduli  one still finds the formal structure of N=2 theory, as in special geometry \cite{deWit:1984pk,Ceresole:1995ca}.

In effective N=1 supergravity obtained by compactification of string theory on CY$_3$ orientifold, one finds that there exists a covariantly holomorphic section from which the K\"ahler potential $K$ is constructed.  The covariantly holomorphic central charge  $Z=e^{K/2}W$ is a symplectic invariant, which transforms under K\"ahler transformations as a phase. 
Therefore $|Z|^2$ and $|DZ|^2$ are both symplectic and K\"ahler invariant. Thus the  geometry of the moduli space, as well as the symplectic invariance of the central charge, remains in the effective N=1 supergravity, which explains the presence of  attractor equations in flux vacua.
Therefore the  potential in (\ref{N1pot}) can be presented in the form
\begin{equation}
   V= |DZ|^2- 3|Z|^2 \ ,
\label{pot}
\end{equation}
where $ Z(t, \bar t, q,p) \equiv  (L^\Lambda
q_\Lambda -
M_\Lambda p^\Lambda)$. Here $p,q$ are integer 3-form fluxes and  $ (L^\Lambda, M_\Lambda) $ is a covariantly holomorphic section for IIB string theory compactified on CY orientifold.

Few more remarks are in order before we start a systematic presentation. 
\begin{itemize}
  \item 
  
The term ``attractor'' in  
Minkowski or dS vacua has a clear meaning in the context of the cosmological evolution 
when {\it the moduli fields start at arbitrary initial conditions and approach the 
minimum of the potential}. 
For AdS vacua the term ``attractor'' may be slightly misleading in the cosmological evolution of the 4d FRW metric towards the $AdS_4$ geometry:
we would find a collapsing universe long before the attractor point is reached, see e.g \cite{Kallosh:2002gg}. Thus we are not replacing the black hole geometry evolving towards the horizon with $AdS_2\times S^2$ geometry by a cosmological evolution towards the $AdS_4$ space. The main interest in supersymmetric AdS flux vacua is related to the fact that they may be uplifted to dS vacua, as in the KKLT construction  \cite{Kachru:2003aw}, which often does not change much the values of the fields fixed in the AdS vacua. The values of the fields in the uplifted dS vacua are indeed attractors during the cosmological evolution. 

\item An additional meaning can be given to the supersymmetric flux vacua attractors if multiple basins of attraction can be discovered for vacua with the same choice of CY space  and the same choice of fluxes. The multiple basins of attraction are known in the  black hole case, see e. g. \cite{Kallosh:1999mz}, \cite{Moore:2004fg}, \cite{Denef:2001xn}. In a situation like that one can look for static supersymmetric domain wall type solutions which interpolate between various attractors, like it was done in 5-dimensional supergravity in \cite{Ceresole:2001wi}.

\item Non-supersymmetric flux vacua  will be also defined here by their attractor equations. The attractors for  any such vacua with non-negative minimal value of the potential will have a clear meaning in the context of the cosmological evolution of the 4d FRW metric towards the de Sitter or Minkowski geometry.

\item 
 The N=1 potential (\ref{N1pot}) based on the GVW superpotential \cite{Gukov:1999ya}
\begin{equation}
W=\int G_3\wedge \Omega
\label{GVW}
\end{equation}
is more general, in principle, than the corresponding potential of the standard gauged supergravity\footnote{The situation is different in more general formulations of extended
supergravities which include antisymmetric tensor fields in four
dimensions \cite{tensors}. Magnetic charges appear together with
electric ones in a
symplectic covariant way, and provide effective masses for the tensor
fields. These magnetic charges may survive in N=1 truncations. I am grateful to G. Dall'Agata, R. D'Auria, B. de Wit and  M. Trigiante for explaining me this new development in gauging supergravities.  } in d=4. This has to do with the fact that only ``electric''  fluxes may appear in the gauged supergravity action.  Therefore in gauged supergravity one may find only an ``electric'' subgroup part of the symplectic symmetry  of the potential (\ref{N1pot}), (\ref{GVW}). In our approach to the attractor problem we will use the case when all fluxes, ``electric'' and ``magnetic,'' may be present, as in Eq. (\ref{GVW}).  
\end{itemize}


\section{ Black Hole Attractors}

\subsection {Black Hole Attractors and Special Geometry} 

Here we give a short overview of attractors and special geometry following \cite{Ferrara:1995ih}-\cite{Ferrara:1996um}, \cite{deWit:1984pk,Ceresole:1995ca}, where we will comment on particular features related to flux vacua. A particularly useful and concise  presentation of the special geometry can be found in \cite{Ceresole:1995ca}.

A special K\"ahler manifold can be defined  by constructing flat
symplectic
bundle of dimension $ 2n+2$ over K\"ahler-Hodge manifold with
symplectic
section defined as
\be
\Pi=(L^\Lambda, M_\Lambda) ,\qquad \Lambda = 0,1,...n\ ,
\label{Pi}\ee
where $(L,M)$ obey the symplectic constraint
$
i(\bar L^\Lambda M_\Lambda - L^\Lambda \bar M_\Lambda)=1
$ and
 $L^\Lambda(t, \bar t) $   and $M_\Lambda(t, \bar t)$ depend on
scalar fields
$t,\bar t$, which are the coordinates of the ``moduli space."
$ L^\Lambda$ and $M_\Lambda$ are {\it covariantly
holomorphic} (with respect to the K\"ahler connection), e.g.
$$
D_{\bar k} L^\Lambda = (\partial_{\bar k} - {1\over 2} K_{\bar
k})L^\Lambda =0
\ ,
$$
 where K is the K\"ahler potential. A Hodge K\"ahler manifold is {\it Special K\"ahler} if there exists a completely symmetric covariantly holomorphic 3-index section $C_{ijk}$ and its covariantly antiholomorphic conjugate $\bar C_{\bar i\bar j\bar k}$ such that the curvature satisfies the ``special geometry'' constraint:
\be
R_{i\bar j l\bar m}= G_{i\bar j}G_{l\bar m}  + G_{i\bar m}G_{l\bar j}- C_{ilp} \bar C_{\bar j \bar m \bar p} G^{p\bar p} \ .
\ee
Symplectic invariant  form of
the K\"ahler
potential can be found  by introducing the {\it holomorphic
section}  $(X^\Lambda (t), G_{\Lambda}(t))$:
$$
L^\Lambda = e^{K/2} X^\Lambda \ , \qquad M_\Lambda = e^{K/2} G_\Lambda\ ,
\qquad
(\partial_{\bar k}
X^\Lambda  = \partial_{\bar k} G_\Lambda=0) \ .
$$
The K\"ahler potential is
$
K = -\ln i( \bar X^\Lambda G_\Lambda -   X^{\Lambda } \bar G_\Lambda).
$
The K\"ahler  metric is given by $G_{k\bar k} = \partial_k
\partial_ {\bar k}
K$. A covariant derivative is defined as $ D_i \Pi=(\partial_i +{1\over 2} K_{,i})\Pi $ and one finds that in special geometry
\be
 D_i D_j \Pi = i C_{ijk} G^{k\bar k} \bar D_{\bar k} \bar \Pi \ .
\label{second}\ee
Finally, from special geometry one finds that there exists a complex
symmetric $(n+1)\times (n+1) $ matrix ${\cal N}_{ \Lambda
\Sigma}$ such  that
\begin{equation}
M_\Lambda = {\cal N}_{ \Lambda \Sigma} L^\Sigma \ , \qquad
{ \rm {Im\,}}{\cal N}_{ \Lambda \Sigma} L^\Lambda  \bar L^\Sigma
=-{1\over 2} \ ,
\qquad
D_{\bar i} \bar  M_\Lambda = {\cal N} _{\Lambda \Sigma} D_{\bar i}  \bar
L^\Sigma \ .
\label{Ncal}
\end{equation}
In the case of N=2 supergravity this matrix is a metric in the vector part of the moduli space depending on scalar fields $t, \bar t$. It is, however, important to realize that we will only use here (in effective N=1 supergravity) the fact that the explicit expression for ${\cal N}$ is defined by the section and the derivative of it over the moduli.

One can introduce a symplectic  charge related to the charge of the graviphoton in 4-dimensional supergravity, it may related to the  integer flux in a compactified internal manifold:
$(
p^\Lambda  ,  q_\Lambda)
$.
Now  we may define a {\it covariantly holomorphic central charge} 
\be
Z(t, \bar t, q,p) \equiv  (L^\Lambda
q_\Lambda -
M_\Lambda p^\Lambda) \ ,
\label{central}\ee 
where $D_{\bar i} Z\equiv (\partial_{\bar i} - {1\over 2} K_{
\bar i}) Z =0$ and $D_i \bar Z\equiv (\partial_{ i} - {1\over 2} K_{
 i}) \bar Z = 0$.
It follows from eq.  (\ref{second}) that in special geometry
\be
D_{ i} D_j  Z= i C_{ijk} G^{k\bar k} \bar D_{\bar k} \bar Z \ .
\ee
One could also define a {\it holomorphic central charge}  
\be
W= e^{-K(t, \bar t)/2} Z(t, \bar t, q,p) \equiv  (X^\Lambda
q_\Lambda -
G_\Lambda p^\Lambda) \ , \qquad \partial_{ \bar i}W=0\ .
\label{superpotential}\ee 
This holomorphic charge $W$ may be associated with  the superpotential in N=1 effective supergravity for the IIB string theory compactified on CY orientifold.
In the generic point of the moduli space there are  two symplectic    invariants homogeneous of degree 2 in electric and magnetic charges:
\begin{equation}
I_1 =  I_1(p,q,t,\bar t)=-{1\over 2} P^t {\cal M}({\cal N}) P\ ,
\qquad
I_2 = I_2(p,q,t,\bar t)=-{1\over 2} P^t {\cal M}({\cal F}) P \ . 
\label{F}
\end{equation}
Here $P=(p,q)$ and ${\cal M}({\cal N})$ is the real symplectic $(2n+2) \times (2n+2)$ matrix
$$
\pmatrix{
{\rm Im} {\cal N} + {\rm Re} {\cal N} {\rm Im} {\cal N}^{-1} {\rm Re} {\cal N}\ &&& - {\rm Re} {\cal N} \,
{\rm Im} {\cal N}^{-1} \cr
-{\rm Im} {\cal N}^{-1}  {\rm Re} {\cal N}  \ &&&{\rm Im} {\cal N}^{-1}  \cr
}
$$
The matrix ${\cal M}({\cal F})$ is an analogous matrix with ${\cal N}$ replaced by ${\cal F}$, where ${\cal F}= \partial_\Lambda G_\Sigma $.
Using the central charge, one can rewrite these two invariants as follows:
\begin{equation}
I_1 =  |Z|^2 + |D_i Z|^2 \ , \qquad
I_2 =  |Z|^2 - |D_i Z|^2 \ .
\label{inv}
\end{equation}

An extremization condition for both symplectic invariants, which specifies the  values of moduli in terms of charges,  is given by 
\be
D_i Z(t, \bar t, q,p) \equiv (\partial_{ i} + {1\over 2} K_{
i}) Z =0 \ , \qquad \bar D_{\bar i} \bar  Z(t, \bar t, q,p) \equiv (\partial_{ \bar i} + {1\over 2} K_{
\bar i}) \bar Z =0 \ .
\label{extrem}\ee 
At this point we find the extremal value of the square of the central charge which is proportional to the black hole area of the horizon
\begin{equation}
 |Z|^2_{cr} (p,q) = {A(p,q)\over 4\pi} \ .
\label{area}
\end{equation}
Equation (\ref{extrem}) is also a requirement of an unbroken supersymmetry.
In terms of the holomorphic charge $W$, the extremization condition is even more familiar:
$$
 D_i W\equiv (\partial_{ i} +  K_{
i}) W =0 \ . $$ 
Note that $X^\Lambda(t)$ are subject to holomorphic redefinitions (sections of a holomorphic line bundle):
$$\label{xx1}
X^\Lambda(t) \to X^\Lambda(t)~e^{-f(t)} \ ,
$$
so that $$\label{xx2}
L^\Lambda(t) \to  L^\Lambda(t)~e^{\bar f(\bar t) -f(t)\over 2}\  .
$$
This occurs because $L^\Lambda = e^{K/2} X^\Lambda$ and $K\rightarrow K+f+\bar f$
under  K\"ahler transformations, so that 
\begin{equation}
 Z(q,p,t, \bar t) \to Z(q,p,t, \bar t)~e^{\bar f(\bar t) -f(t)\over 2} \ .
\label{xx3}
\end{equation}
However,  $|Z|$ is both symplectic and K\"ahler gauge invariant, this is why the connection drops  and $D_i Z = 0$ ($D_{\bar i}Z \equiv 0$) means that $\partial_i |Z| = 0$.  This explains why at the attractor point, $D_i Z=0$, the square of the central charge  does not depend on moduli, only on charges:
\be
 |Z( t(p,q), \bar t(p,q),q,p)|^2 \ .
\label{attractorZ}
\ee
 The critical point of the black hole mass $M_{bh}^2= |Z( t(p,q), \bar t(p,q),q,p)|^2$ at $D_i Z=0$ can be also presented in the form of the  attractor equations
\begin{equation}
 \left (\matrix{
p^\Lambda\cr
q_\Lambda\cr
}\right )= i \left(\matrix{
\bar Z L^\Lambda- Z\bar L^\Lambda\cr
 \bar Z M_\Lambda- Z \bar M_\Lambda\cr
}\right ) \ .
\label{stab}
\end{equation}
From  these equations it follows that $(p,q)$  determine the sections up to a
(K\"ahler) gauge transformation. The moduli at the fixed point depend on ratios of charges since the equations  are homogeneous in $p,\,q$. According to special geometry the supersymmetric attractor point describes {\it the minimum of the black hole mass} since the only non-vanishing second derivative is strictly positive-definite \cite{Ferrara:1997tw}
\be
\partial_i \partial_{\bar k} |Z(t(p,q), \bar t    (p,q),q,p)|_{fix} = {1\over 2}G_{i\bar k} |Z|_{fix} \ .
\ee
Here we consider only black holes with a regular horizon, which means that $|Z|_{fix}\neq 0$ and the metric of the moduli space $G_{i\bar k}$ must be positive definite. 

For completeness we present here the derivation of these equations following \cite{Ferrara:1996dd}. It will be clear that the special geometry of the moduli space plays an essential role in this derivation.  Thus if the flux vacua satisfy the condition $D_{ i} D_j  Z=0$, to be consistent with special geometry where
\be
D_{ i} D_j  Z= i C_{ijk} G^{k\bar k} \bar D_{\bar k} \bar Z =0 \qquad \rm at \qquad \bar D_{\bar k} \bar Z=0 \ ,
\label{spec}\ee
we may expect  that the attractor equations for flux vacua have an analogous form. In more general case, terms proportional to $D_{ i} D_j  Z$ may be needed.

Start with
\begin{equation}
D_{\bar i}\bar Z   =  D_{\bar i}\bar L^\Lambda q_\Lambda -  D_{\bar i}\bar M_\Lambda p^\Lambda = 0 \ .
\label{dif}
\end{equation}
Now replace $D_{\bar i}\bar M_\Lambda$ by ${\cal N}_{\Lambda\Sigma}D_{\bar i}\bar L^{\Sigma}$, according to eq. (\ref{Ncal}).
By contracting with $D_iL^\Sigma G^{i\bar i}$ and using the identity
$
D_iL^\Sigma G^{i\bar i} D_{\bar i}\bar L^\Lambda = - {\textstyle {1\over 2}} {{\rm Im}} {({\cal N}^{-1})}^{\Sigma\Lambda} - \bar L^{\Sigma}L^\Lambda 
$
we get 
\begin{equation}\label{u2}
2Z \bar L^\Sigma = i p ^\Sigma - {{\rm Im}} {({\cal N}^{-1})}^{\Sigma\Lambda}\, q_\Lambda + {{\rm Im}}  {({\cal N}^{-1})}^{\Sigma\Gamma}\  {{\rm Re}}   {\cal N}_{\Gamma\Delta}\  p^\Delta  \ ,
\end{equation}
from which it follows that
\begin{equation}\label{u3}
  p^{\Sigma}  =  2i\bar Z  L^\Sigma   - i  ({{\rm Im}}  {\cal N}^{-1}\  {{\rm Re}}  {\cal N}\, p  +  {\rm Im} {\cal N}^{-1} \, q )^\Sigma \ ,
\end{equation}
\begin{equation}\label{u4}
q_{\Sigma} = 2i\bar Z  M_\Sigma     -i  ({{\rm Im}}  {\cal N} \, p  + {{\rm Re}}  {\cal N}\    {{\rm Im}} {{\cal N}^{-1} }\  {{\rm Re}}\,  {\cal N} \, p - {{\rm Re}} {\cal N}\  {{\rm Im}}  { {\cal N}^{-1} }\, q )_\Sigma\ .
\end{equation}
Now we see that it is really important to have a symplectic charge $(p^\Lambda, q_\Lambda)$ which is real and moduli independent: note that we have differentiated  in eq. (\ref{dif}) over the moduli inside $\bar Z$ only the section  $\bar L^\Lambda, \bar M_\Lambda$,  not the flux $p,q$. In such case we find from eqs. (\ref{u3}) and (\ref{u4}) the attractor eqs. (\ref{stab}).

This detailed derivation of the attractor equations  explains why the use of the $G_3=F_3-\tau H_3$ flux in type IIB theory  obscures the derivation of the attractor equations for the flux vacua. It also suggests that one should try to rewrite the potential for N=1 effective supergravity in IIB theory compactified on a CY orientifold in a form suitable for the application of attractor equations: {\it one should  use only the quantized fluxes not mixed with moduli, and one should push all dependence on all moduli into the properly defined symplectic section.}


\subsection{Black Hole Attractors and Calabi-Yau Threefolds in IIB String  Theory}

In case of black holes in IIB string  theory compactified on Calabi-Yau$_3$-folds we have a particular choice of special geometry in effective 4-dimensional supergravity. In such case one can derive the black hole attractor equations in the spirit of  \cite{Strominger:1996kf}, \cite{Moore:2004fg}. We will do it here in a form  which will be useful for a consequent comparison of flux vacua with black hole attractors and we will use here a Hodge decomposition of the kind used in \cite{Denef:2004ze} for the 4-form in flux vacua.

We start by defining 
the holomorphic symplectic basis on CY  such that
\be
  X^a(x)= \int_{A^a} \Omega \ , \qquad G_a(x)= \int_{B_a} \Omega \ ,
\label{holomorphic} 
\ee
and 
\begin{equation}
  \Omega_3= X^a(x)\, \alpha_a - G_{a}(x)\, \beta^a \ ,
\end{equation}
where
$$
  \int_{CY} \alpha_a\wedge \beta^b= \delta_a{}^b \ .
$$
Here $ \Omega(x)$ is a  holomorphic 3-form on a Calabi-Yau space depending on the complex structure moduli $x$, and the K\"ahler potential of CY is given by
\begin{equation}
 K(x, \bar x)= - \ln [ i\int_{CY}  \Omega(x)\wedge \bar \Omega(\bar x)]= - \ln i[\bar X^a G_a-X^a\bar G_a]  \ .
\label{K}
\end{equation} 
Assume that the manifold admits a real 3-form flux, 
\begin{equation}
   H_3= p^a \,\alpha_a - q_{a}\, \beta^a \ ,
\label{H3}
\end{equation}
In addition to a holomorphic 3- form it will be useful to introduce a covariantly holomorphic 3-form 
\be
\hat \Omega= e^{K\over 2}\Omega= L^a(x)\, \alpha_a - M_{a}(x)\, \beta^a \ 
\ee
such that 
\begin{equation}
1=   i\int_{CY}  \hat \Omega \wedge \hat {\bar \Omega}=  i[\bar L^a M_a-L^a\bar M_a]  \ .
\label{1}
\end{equation} 
We now define a central charge as follows
\be
Z= \int_{CY} H_3\wedge \hat \Omega = L^a(x)\, \alpha_a - M_{a}(x)\, \beta^a \  .
\ee
As suggested by Hodge decomposition, a real 3-form flux can be presented as a combination of all possible independent covariantly holomorphic and anti-holomorphic forms 
\be
H_3= i \, [\bar Z \, \hat \Omega_3 -G^{i\bar i}(\bar D_{\bar i} \bar Z)\,  D_i \hat \Omega_3 ] +c.c.
\label{Hodge}\ee 
An analogous equation has been derived and studied in \cite{Ferrara:1990dp}. 
Note that the term $D_A D_B \hat \Omega_3 $ can be expressed via $\bar D^A \hat {\bar \Omega}_3$ on Calabi-Yau 3-fold  as in special geometry equation (\ref{second}).
\

The second term in (\ref{Hodge}) vanishes at the supersymmetric minimum $DZ=0$. Thus the only terms describing the supersymmetric black hole attractor equation remaining in eq.  (\ref{Hodge}) are
\be
H_3= i [\, \bar Z \, \hat \Omega_3 - i \,  Z \, \hat {\bar \Omega}_3]_{fix} = 2 \rm Im (Z \, \hat {\bar \Omega}_3)|_{DZ=0} \ .
\label{crit}\ee 
With account of all definitions above one can see that these equations are exactly the equations (\ref{stab}) derived in the context of special geometry in the framework of 4-dimensional supergravity. In fact, equations in this section present a particular example of equations in (\ref{stab}) which may be associated with a class of manifolds of special geometry more general than Calabi-Yau.

We went into a lengthy details of the black hole attractor equation (\ref{crit}) to be able to explain the common features and differences with flux vacua.


\subsection{New Non-supersymmetric Black Hole Attractor Equations}

The non-supersymmetric black hole attractors were introduced in \cite{Ferrara:1997tw} and studied recently in \cite{Goldstein:2005hq}. The story boils down to the  extremization of the ``black hole potential''
\be
V_{BH}= I_1 =  |Z|^2 + |D_i Z|^2  \ .
\ee
At the non-supersymmetric critical point $D_i Z\neq 0$ but
\be
\partial_i V_{BH}= \partial_{\bar k} V_{BH} =0 \ .
\ee
The attractor equations take the following form
\be
H_3= 2 {\rm Im} \left [  \,  Z \, \bar {\hat \Omega}_3 -  G^{i\bar i}  D^{ i}  Z 
\bar D_{\bar i} \hat {\bar \Omega}_3 \right ]_{ \partial V_{BH} =0}
\label{nonsusyBH}\ee  
This equation can be made explicit using the equation $\partial V_{BH} =0$ in the form  $2(D_i Z) \bar Z + i C_{ijk}G^{j\bar m} G^{k\bar k} \bar D_{\bar m} \bar Z  \bar D_{\bar k} \bar Z =0$, as shown  in \cite{Ferrara:1997tw}.

\section{ Type IIB Flux Vacua as Attractors, $M_{AB}=0$ case} 

Consider  a compactification of IIB string theory on some CY 3-fold with O3/O7 planes. We refer the reader to a nice recent review of the related topics in \cite{Grana:2005jc}. The 3-cycles come in pairs (A, B) so that (with $(2\pi)^2 \alpha'=1$) the RR 3-form $F_3$ and the NS 3-form $H_3$ are 
\begin{equation}
  F_3= p_f^a \,\alpha_a - q_{af} \,\beta^a \ , \qquad  H_3= p_h^a \,\alpha_a - q_{ah}\, \beta^a \ ,
\label{H3}
\end{equation}
 where 
\begin{equation}
  \int_{CY} \alpha_a\wedge \beta^b= \delta_a{}^b \ .
\label{norm}
\end{equation}
The holomorphic symplectic basis on CY is the same as in the previous section.

This standard description of the compactified IIB string theory is supplemented by the axion-dilaton moduli as follows. From the $SL(2,Z)$ doublet of fluxes, $F$ and $H$, one forms a complex 3-form $G_3= F_3-\tau H_3$, and the K\"ahler potential has an additional part, so that the total K is
\begin{equation}
  K(\tau, \bar \tau, x, \bar x)=- \ln [-i (\tau-\bar \tau)]- \ln [ i\int  \Omega(x)\wedge \bar \Omega(\bar x)] \ .
\label{dialtonkahler}
\end{equation}
The superpotential is given by
\begin{equation}
  W= \int_{CY} G_3(\tau)\wedge \Omega(x)= (p_f^a -\tau p_h^a )G_a - (q_{fa} -\tau q_{ha})X^a \ .
\label{W}
\end{equation}
It is this particular dependence on the axion-dilaton that makes the symplectic structure of the flux vacua obscure.

Let us instead make the $SL(2, Z)$ symmetry of the type IIB theory manifest, so that it extends the manifest symplectic symmetry of the CY space. The first hint comes from rewriting the superpotential as follows: 
\begin{equation}
  W= \int_{CY} G_3(\tau)\wedge \Omega(x)= p_f^a G_a- q_{fa}X^a+
  p_h^a (-\tau  G_a)   -  q_{ha}(- \tau X^a) \ .
\label{W1}
\end{equation}
Let us now introduce  the ``central charge'' in a form useful for the attractor equation: 
\begin{equation}
Z=e^{K(x, \bar x, \tau, \bar \tau)\over 2}  W (\tau, x)= e^{K(x, \bar x, \tau, \bar \tau)\over 2}\int F_3\wedge \Omega + H_3\wedge (-\tau \Omega) \ .
\label{Z}
\end{equation}
Thus we have a 3-form flux $SL(2,Z)$ doublet
\begin{equation}
 F= (F_3, \, H_3) \ ,
\label{flux doublet}
\end{equation}
and a symplectic section, which is also a doublet,
\begin{equation}
\Xi= \left (\matrix{
\Xi_1(\tau, x)\cr
\Xi_2(\tau, x)\cr
}\right )
  = \left (\matrix{
\Omega(x)\cr
-\tau \Omega(x)\cr
}\right ) \ .
\label{sectiondoublet}
\end{equation}
The total K\"ahler potential is now given by
\begin{eqnarray}
K(x, \bar x, \tau, \bar \tau )&=& - \ln \left[ \int [ \tau \Omega(x)\wedge \bar \Omega(\bar x)-  \Omega(x)\wedge \bar \tau \bar \Omega(\bar x)]\right]\nonumber\\
&=& -\ln \left[\int (\Xi_1\wedge \bar  \Xi_2 -  \Xi_2\wedge \bar \Xi_1)\right] \ .
\label{Kahler}
\end{eqnarray}
The central charge is
\begin{equation}
Z=e^{K\over 2}  W = e^{K\over 2} \int F\wedge \Xi \ .
\label{newZ}
\end{equation}


\subsection{$SL(2,Z)$ symmetry}

Under $SL(2,Z)$ transformations with
\begin{equation}
   \tau'= {a\tau +b\over c\tau +d}
\label{R}
\end{equation}
the flux doublet transforms as 
\begin{equation}
  \left (\matrix{
F_3\cr
H_3\cr
}\right )'= R   \left (\matrix{
F_3\cr
H_3\cr
}\right )  \ , \qquad R= \left (\matrix{
a & b\cr
c& d\cr
}\right )\ .
\label{fluxtransf}
\end{equation}
The central charge, defined in eqs. (\ref{Z}), (\ref{newZ}) transforms with the phase
\begin{equation}
  Z' = e^{-i Arg(c\bar \tau+d)} Z \ ,
\label{Ztransf}
\end{equation}
as was shown in studies of axion-dilaton black holes in \cite{Kallosh:1993yg}.
This is clearly compensated by the K\"ahler transformation of the type (\ref{xx3}).
This concludes the derivation of the  special geometry in the moduli space $z=(\tau, x)$ of the axion-dilaton $\tau$ and the complex structure fields $x$ of the CY$_3$. The attractor equations for the flux vacua of type IIB string theory on CY$_3$ orientifold can be guessed by analogy with black hole attractor equations, at least for the case that the special geometry rule is satisfied, i. e. at the fixed point $DZ=0$ with additional requirement that $M_{AB}= D_A D_B Z=0$
\begin{equation}
  \left (\matrix{
p_h^a\cr
\cr
q_{ha}\cr
\cr
p_f^a\cr
\cr
q_{fa}\cr
}\right )=   e^{K} \left (\matrix{
 \, \bar W   X^a + \; W \bar X^a\cr
\cr
 \, \bar W   G_a+\; W \bar G_a \cr
\cr
 \tau \bar W   X^a+\bar \tau   W X^a \cr
\cr
 \tau \bar W   G_a+\bar \tau  W \bar G_a\cr
}\right )_{fix}
\label{IIBmassless}
\end{equation}
Here the right hand side has explicit dependence on the universal axion-dilaton and a generic dependence on the complex structure moduli of an arbitrary CY$_3$. All moduli in the rhs. of this equation take fixed values defined by fluxes in the lhs. of the attractor equation.  


\subsection{Simplified notation}

We may also introduce the form notation to simplify the axion-dilaton dependence.
Let  
\begin{equation}
  F_4= - \alpha\wedge F_3 +\beta \wedge H_3 \ , \qquad  \int_{T^2}\alpha\wedge \beta =1 \ .
\label{torus}
\end{equation}
The complex structure of the auxiliary torus is $\omega= \beta - \tau \alpha$. We can define the holomorphic 4-form
\begin{equation}
\Omega\equiv \Omega_{(0,4)}(z)= \Omega_{(0,3)}(x)\wedge\omega(\tau)= \Omega_{(0,3)}\wedge \beta - \tau \Omega_{(0,3)} \wedge \alpha \ .
\label{4period}
\end{equation}
The total K\"ahler potential is now
\begin{equation}
 K (t, \bar t)=  -\ln  \int_{X_3\times T^2}\Omega_{(0,4)}\wedge \bar \Omega_{(4,0)} \ .
\label{Ktorus}
\end{equation}
The covariantly holomorphic central charge and the superpotential are
\begin{equation}
  Z= e^{K\over 2} \int_{X_3\times T^2} F_4\wedge \Omega_{(0,4)} \ ,  \qquad W= \int_{X_3\times T^2} F_4 \wedge \Omega_{(0,4)} \ .
\label{central2}
\end{equation}
The flux attractor equations take a  remarkably simple form
\begin{equation}
F_4= [e^{K} (\bar W \Omega + W \bar \Omega)]_{fix}= 2 e^{K} \rm Re \, ( W \, \bar \Omega_{(4,0)})|_{fix} \ .
\label{attractorsimple}
\end{equation}
We may also present it using the central charge and covariantly holomorphic form $\hat \Omega_4 = e^{K\over 2}\Omega_{(0,4)}$
\begin{equation}
F_4=  [\bar Z \hat \Omega + Z \hat {\bar \Omega}]_{fix}= 2  \rm Re \, ( Z \, \hat {\bar  \Omega})|_{fix} \ .
\label{attractorsimple1}
\end{equation}
This is a condensed form of equation (\ref{IIBmassless}).
The flux vacua attractor equation (\ref{attractorsimple1}) is almost the same equation as the black hole attractor equation. The difference is that in black hole case we have $\rm Im (Z\hat {\bar \Omega}_3)$ defining the 3-form flux whereas in flux vacua case we have $\rm Re (Z\hat {\bar \Omega}_4)$ defining the effective 4-form flux.

The difference is easy to explain. Let us contract both sides of the flux vacua attractor eq. (\ref{attractorsimple1}) with $\hat \Omega_4$. We find 
\be
 Z=  \int_{X_4} F_4\wedge \hat \Omega_{4}= Z \int_{X_4} \hat {\bar \Omega}_4 \wedge \hat \Omega_{4}= Z
\ee 
Here we took into account that $ \int_{X_4} \hat {\bar \Omega}_4 \wedge \hat \Omega_{4}=1$.
Now let us contract both sides of black hole attractor eq. (\ref{crit})  with $\hat \Omega_3$
\be
Z= \int_{X_3}  H_3 \wedge \hat \Omega_3 =   -i Z \int_{X_3}  \hat {\bar \Omega}_3 \wedge \hat \Omega_3  =Z
\ee
Here we took into account that $\int_{X_3}  \hat {\bar \Omega}_3 \wedge \hat \Omega_3 =i$ in agreement with equation (\ref{1}). 

\subsection{Example of M-theory on K3$\times$K3}
As an example, we will consider here 
stabilization of the complex structure moduli in M-theory on K3$\times$K3 \cite{Dasgupta:1999ss,Aspinwall:2005ad,Moore:2004fg}, which is also related to stabilization of the complex structure in IIB string theory on K3$\times T^2\over Z_2$. For the purpose of relating flux vacua with black hole attractors, we will slightly modify the procedure used in \cite{Aspinwall:2005ad}, where we were interested in stabilizing all moduli in this model. For this purpose we  made a choice of the 4-form flux as a $(2,2)$ form. This is a choice when the superpotential defined by eq. (\ref{central2}) vanishes. Here we will fix the complex structure moduli by fluxes in such a way that the superpotential does not vanish: we will choose the real 4-form flux as follows
\begin{equation}
  F_4= \bar c \,
  \Omega^1\wedge \Omega^2+ c \,\bar \Omega^1\wedge \bar \Omega^2 \ .
\label{4flux}
\end{equation}
Here $\Omega^i$,  $i=1,2$ are holomorphic 2-forms on each of K3 at the attractor point.
The superpotential can be calculated, and we find 
\begin{equation}
   W= \int_{K3\times K3}  c \bar \Omega^1\wedge \bar \Omega^2 \wedge \Omega^1\wedge\Omega^2 =  c \, e^{-K}\ , \qquad \Rightarrow \qquad c=Z
\label{super}
\end{equation}
This proves that we may now rewrite the 4-form flux as follows
\begin{equation}
  F_4= 2 e^K \rm Re (\bar W \,
  \Omega^1\wedge \Omega^2) = 2\rm Re (\bar Z \,
  \hat \Omega^1\wedge \hat \Omega^2)  \ .
\label{4fluxattr}
\end{equation}
Since in this example $\Omega_{(0.4)}= \Omega^1\wedge \Omega^2$, we  confirm for this example the   flux vacua attractor formula (\ref{attractorsimple}), (\ref{attractorsimple1}) derived in the general case for type IIB string theory compactified on a generic CY orientifold.

We may now take $\Omega_i, \, i=1,2$ proportional to $ p_i+\tau_i q_i$, as suggested in \cite{Aspinwall:2005ad,Moore:2004fg}. 
Two complex numbers $\tau_j$ for this choice of the holomorphic form  are fixed by the condition
$\Omega_j^2=0$ to be
$
  \tau_j = (-p_j.q_j + i\sqrt{\det Q_j})/{q_j^2}$. Each of the attractive K3 surfaces is now described up to SL(2,Z) equivalence class by a matrix \cite{Aspinwall:2005ad}
\begin{equation}
 Q_j= \left (\matrix{
p_j^2&p_j.q_j\cr
\cr
p_j.q_j&q_j^2\cr
}\right )
\end{equation}\label{qi2}
 (no summation over $j$).                     
In  the definition of the attractive K3 surface  in \cite{Moore:2004fg} it was pointed out that  ${A\over 4\pi}=|Z|_{fix}^2= \sqrt{ Det \, Q}$
is an area of the unit cell in the transcendental lattice $T_S$ of the K3 surface. 
As we will see now, our example  helps to reveal another interesting relation between black holes and flux vacua: a relation between the black hole entropy and the transcendental lattice  $T_S$ on the attractive K3.

Indeed, 
the area of the horizon of the supersymmetric black holes in ${SU(1,1)\over U(1)} \times {SO(2,n)\over SO(2)\times SO(n)}$ symmetric
manifold was established in \cite{Kallosh:1996tf}:
\begin{equation}
  {A\over 4\pi}= |Z|_{fix}^2= \sqrt{ Det \, Q}=  (p^2 q^2- (p\cdot q)^2)^{1/2} \ .
\label{blackhole}
\end{equation}
This result was found by solving explicitly the supersymmetric black hole attractor equations in this theory and stabilizing all moduli of the ${SU(1,1)\over U(1)} \times {SO(2,n)\over SO(2)\times SO(n)}$ coset space in terms of the black hole charges. Now   one finds   the same  value of $|Z|_{fix}^2$  in the context of flux vacua for each K3.  We leave the detailed study of this issue for the future publication.

Note that Eq. (\ref{attractorsimple}), which in detailed form is given in Eq. (\ref{IIBmassless}), was derived either using the tools of special geometry or using the tools of algebraic geometry such as Hodge-decomposition  of the space of allowed 4-form fluxes.  It is valid for generic CY$_3$. As we have shown in particular example of K3$\times {T^2/Z_2}$ and its M-theory version, the coefficient of proportionality between fluxes and period matrix is linear in the critical value of the superpotential (central charge).


\section{New Attractors: General type IIB Flux Vacua, $M_{AB}\neq0$ case}

Here again we study IIB string theory and we take $X_4$ to be an elliptically fibered CY 4-fold ${ T^2\times Y\over Z_2}$. $F_4= - \alpha\wedge F_3 +\beta \wedge H_3  $ is a real flux 4-form representing a doublet of RR and NS forms of type IIB string theory. $\hat \Omega_4$ is a covariantly holomorphic 4-form on $X$ 
\begin{equation}
  D_{\bar a} \hat \Omega_4= (\partial_{\bar a} -{1\over 2} K_{, \bar a}) \hat \Omega_4 = 0 \ ,\qquad a=(0,i) \ ,
\label{covholom}
\end{equation}
\begin{equation}
  \int_X \hat \Omega_4 \wedge \hat {\bar \Omega}_4 = 1  \ .
\label{norm}
\end{equation}
$\hat \Omega_4(t, \bar t)$  is related to the holomorphic 4-form $\Omega_{(4.0)} (t)$, $\hat \Omega_4= e^{K\over 2}\Omega_{(4.0)} (t)$ and 
\begin{equation}
  \int_X   \Omega_{(4.0)} (t^a) \wedge  \bar \Omega_{(0,4)}(\bar t^{\bar a})= e^{-K} \ .
\label{norm}
\end{equation}
In our particular case when $X={ T^2\times Y\over Z_2}$, \,   $\hat \Omega_4=\hat \Omega_1\wedge \hat \Omega_3$ where  $\hat \Omega_1$ is a covariantly holomorphic 1-form on $T^2$ 
\begin{equation}
\hat \Omega_1= e^{K_1\over 2}(\beta-\tau \alpha) \ ,  \qquad K_1= -\ln i(\bar \tau- \tau) \ , \qquad \tau=t^0\ .
\label{1form}
\end{equation}
and $\Omega_3= e^{{K_3(t^i, \bar t^i)\over 2} }\Omega_{(3,0)}(t^i)$ is a covariantly holomorphic  3-form on  a generic CY orientifold with the corresponding holomorphic form $\Omega_{(3,0)}(t^i)$ and $i=1, ..., n$.
Eq. (2.20) in \cite{Denef:2004ze} proposes the following decomposition of the real 4-form flux into various possible (m,n) forms on a 4-fold $X$.
\begin{equation}
  F_4= \bar Z \hat \Omega- \bar D^{ A}\bar  Z \, D_A \hat \Omega + \bar D^{\underline 0 I}\bar Z \,  D_{\underline 0 I} \hat \Omega + c.c.
\label{G4}
\end{equation}
The capital index $A, \bar A$ is the index associated with the orthonormal frame $e^a_A$ such that $\delta_{A\bar B}= e^a_A g_{a\bar b} e^{\bar b}_{\bar B}$ and  $A=(\underline 0, I)$.

We will be interested in supersymmetric vacua here, which are also critical points of the  N=1 F-term potential. Keeping the second term in eq. (\ref{G4}) may help us to understand the attractor behaviour of non-supersymmetric black holes, studied  in \cite{Ferrara:1997tw,Goldstein:2005hq}. However, here we will focus on supersymmetric vacua (\ref{susy}): Minkowski with $Z=0$ and AdS with $Z\neq 0$.

We rewrite eq. (\ref{G4}) in a form in which it is obviously an attractor equation, defining a combination of moduli via some integer charges. First we find that 
\begin{equation}
  F_4=\left(\bar M_{3/2}\hat \Omega_1\wedge \hat \Omega_3 - \bar M^{\underline 0 I}\hat { \bar \Omega}_1 \wedge \chi_I\right) +c.c. 
\label{attr}
\end{equation}
Here  
\begin{equation}
\chi_I\equiv D_I \hat \Omega_3= e_I^i (\partial_i+ {1\over 2}K_{ ,i}) \,\hat \Omega_3
\label{chi}
\end{equation}
and \begin{equation}
  M^{\underline 0 I} = D^{\underline 0}D^I Z
\label{mix}[t , \bar t ; p, q]\equiv Z^{\underline 0 I} \ .
\end{equation}
We have also used the fact that $D_{\underline 0}\hat \Omega_1=\hat {\bar \Omega_1}$.
Now we may rewrite these equations in a form in which it is easy to recognize them as generalized attractor equations. 
\begin{equation}
  \left (\matrix{
p_h^a\cr
\cr
q_{ha}\cr
\cr
p_f^a\cr
\cr
q_{fa}\cr
}\right )=    \left (\matrix{
 \, \bar Z   L^a+ \; Z \bar L^a\cr
\cr
 \, \bar Z   M_a+ \; Z \bar M_a \cr
\cr
 \tau \bar Z   L^a+\bar \tau   Z L^a \cr
\cr
 \tau \bar Z   M_a+\bar \tau  Z \bar M_a\cr
}\right )_{DZ=0}+
  \left (\matrix{
 \, \bar Z^{\underline 0I}   D_I L^a+\; Z^{\underline 0I} \bar D_I \bar L^a\cr
\cr
 \, \bar Z^{\underline 0I}  D_I M_a+ \; Z^{\underline 0I} \bar D_I \bar M_a \cr
\cr
\bar \tau \bar Z^{\underline 0I} D_I   L^a+ \tau   Z^{\underline 0I} \bar D_I L^a \cr
\cr
\bar \tau \bar Z^{\underline 0I}  D_I M_a+ \tau  Z^{\underline 0I} \bar D_I \bar M _a\cr
}\right )_{DZ=0}
\label{IIB}
\end{equation}
Here $p_f, q_f$ are  magnetic and electric charges associated with the RR 3-form flux $F_3$ and $p_h, q_h$ are magnetic and electric charges associated with the NS 3-form flux. The special geometry of the CY$_3$ manifold is codified in the  covariantly-holomorphic section $(L^a, M_a)$.

All  supersymmetric flux vacua for CY$_3$ orientifold with vanishing  or non-vanishing $W$ are expected to satisfy the attractor equations (\ref{IIB}). We may rewrite them in a slightly shorter form using $f=(p^a_f, q_{af})$ and $h=(p^a_h, q_{ah})$ and $\Pi= ( L^a,  M_a)$

\begin{equation}
  \left (\matrix{
h\cr
\cr
f\cr
}\right )=   \left (\matrix{
 \, 2 \rm Re \, (\bar Z  \, \Pi) \cr
\cr
 \,   2 \rm Re ( \bar Z \, \tau  \Pi )\cr
}\right )_{DZ=0}+ 
  \left (\matrix{
 \, 2 \rm Re (\bar Z^{\underline 0I} \,  D_I \, \Pi)\cr
\cr
 \, 2 \rm Re ( \bar Z^{\underline 0I} \,\bar \tau \, D_I \Pi ) \cr
}\right )_{DZ=0}
\label{IIBshort}
\end{equation}
Even if the first term in the right hand side of eq. (\ref{IIB}),  (\ref{IIBshort}) vanishes, i. e. if $Z_{fix}=0$, the second term gives a solution of the attractor equations and allows to find a critical value of the axion-dilaton and complex structure moduli as functions of all fluxes $p,q$.

It may be useful to explain why in the black hole case the second term in the right hand side of the  attractor eqs.(\ref{IIB}), (\ref{IIBshort}) is absent. Special geometry rules related to CY$_3$ manifold require that 
\begin{equation}
M_{IJ}=  D_I D_J Z = i C_{IJK}\bar D^{K}\bar Z \ .
\label{mass}
\end{equation}
At the supersymmetric critical point $DZ=0$,  $M_{IJ}=0$. Meanwhile, in flux vacua in type IIB string theory the complex structure modulino mass matrix is not vanishing at the supersymmetric critical point as it is proportional to the dilatino-modulino mixing \cite{Denef:2004ze}, 
\begin{equation}
  D_I D_J Z=  C_{\underline 0 IJK}\bar D^{\underline 0} \bar D^I \bar Z = C_{ IJK} \bar M^{\underline 0 K} \ .
\label{modulino}
\end{equation}
The point is that in the case of N=1 flux vacua in type IIB string theory the corresponding geometry is only partially related to special geometry of CY$_3$. As a result,  the second derivative of the central charge has a contribution from the dilatino-modulino mixing which does not vanish at the critical point, in general.  $ D_{\underline 0 I} Z$  cannot be expressed via the first derivative of the central charge since we have the product geometry of the  torus with CY and the chiral fermion mass term does not have to vanish at the supersymmetric vacua. Its presence   provides the new term in the attractor equation as shown in eq. (\ref{IIB}).

One can summarise this in the following way. Equation (\ref{IIB}), (\ref{IIBshort}) is a general flux attractor equation for type IIB string theory on CY$_3$ orientifold. One can study its general solutions taking into account both terms in its right hand side. On the other hand, one may try first to find particular solutions for which the  dilatino-modulino mixing  vanishes at the critical point. In this case the second term in equation (\ref{IIBshort}) disappears, and it acquires the form given in  \cite{Kallosh:2005bj}: 
\begin{equation}
  \left (\matrix{
h\cr
\cr
f\cr
}\right )=   \left (\matrix{
 \, 2 \rm Re \, (\bar Z  \, \Pi) \cr
\cr
 \,   2 \rm Re ( \bar Z \, \tau  \Pi )\cr
}\right )_{DZ=0}
\label{IIBshort1}
\end{equation}
In this  case one has a full analogy between flux vacua and black hole attractors, as discussed in  \cite{Kallosh:2005bj}, apart from the replacement of ``$\rm Im$'' in black hole case   by ``$\rm Re$'' in flux vacua case. In particular, the square of the gravitino mass plays the  role in the theory of flux attractors analogous to  the entropy in the theory of the black hole attractors.

However, this class of solutions does not describe possible solutions for which the superpotential and the gravitino mass $M_{3/2} = Z$ vanish at the critical point.  To find such solutions one can leave only the second term in equations (\ref{IIB}), (\ref{IIBshort}): 
\begin{equation}
  \left (\matrix{
h\cr
\cr
f\cr
}\right )=   
  \left (\matrix{
 \, 2 \rm Re (\bar Z^{\underline 0I} \,  D_I \, \Pi)\cr
\cr
 \, 2 \rm Re ( \bar Z^{\underline 0I} \,\bar \tau \, D_I \Pi ) \cr
}\right )_{DZ=0}
\label{IIBshort2}
\end{equation}
In a general case, one should use eq. (\ref{IIBshort}) with both terms on the right hand side present.

In the black hole case the number of differential equations $D_a Z=0$ is equal to $2n+2$, where $n$ is the number of special coordinates in special geometry. The number of black hole attractor equations (\ref{stab})
is also equal to $2n+2$. The counting of new attractor equations for a set of flux vacua goes as follows:
The original complex equation specifying the vacua is $D_A Z=0$, which consists of a set of $2n+2$ independent equations but the number of equations in the attractor form in (\ref{IIB}) is  $2(2n+2)$. This leaves us with a puzzle which may be resolved as follows. An explicit calculation of all intersection numbers in elliptically fibered CY 4-fold ${ T^2\times Y\over Z_2}$ has been performed in \cite{Denef:2004ze}. The model has some number of relations between the derivatives of various order acting on the covariantly holomorphic central charge, see eqs. (2.21)-(2.27) in \cite{Denef:2004ze}. With account of these relations, equation for the real integer set of fluxes in $G_4$ has in the right hand side  all possible independent  forms available in this model. Therefore we have in our model for IIB compactification on CY$_3$ orientifold  only one new term in the right hand side of the equation for $G_4$ in addition to the first term, which is a familiar black hole attractor type term. 

In other models of M/string theory where the simplifying relations are partially or totally absent we may still have an attractor type equation relating the values of the integer fluxes to some combination of moduli and fluxes. For example, in M-theory we may have an attractor relation of the kind
\begin{equation}
  F_4= \bar Z \Omega_4 + Z \bar \Omega_4 +\cdots
\label{Mtheory}
\end{equation}
Here $\Omega_4$ is a covariantly holomorphic 4-form on a 4-fold $X$ such that $\int _X \Omega_4\wedge \bar \Omega_4=1$. (We omit $\hat{}$ over $\Omega$ in these examples). All terms with dots in eq. (\ref{Mtheory}) are orthogonal to $\Omega_4$ and to $\bar \Omega_4$ and  the central charge is given by
\begin{equation}
  Z= \int_X G_4 \wedge \Omega_4 = \int_X Z \, \bar \Omega_4\wedge \Omega_4=Z \ .
\label{MtheoryZ}
\end{equation}
In particular, for $X=K3\times K3$ we may have the following 2 possibilities. One   can take
\begin{equation}
  F_4= \bar Z \Omega_2^1\wedge  \Omega_2^2 + Z \bar \Omega_2^1\wedge  \bar \Omega_2^2 \ ,
\label{R}
\end{equation}
as suggested in \cite{Kallosh:2005bj}.
Here $\Omega_2^1$ and $ \Omega_2^2$ are the covariantly holomorphic 2-forms on each of $K3$. This looks similar to a  black hole attractor equation corresponding to AdS vacua with $V_{cr} = -3 |Z_{cr}|^2$.
Another possibility, suggested in \cite{Tripathy:2002qw} for type IIB theory compactified on K3$\times {T^2\over Z_2}$\,,
 is to take the case with $Z_{cr}=0$. In condensed notation with covariantly holomorphic forms on $K3$, on torus and on auxiliary torus we find
\begin{equation}
  F_4= c \, \bar \Omega_2 (K3) \wedge  \Omega_1(\tau, \bar \tau)\wedge \Omega_1 (\phi, \bar \phi)+cc \ .
\label{TT}
\end{equation}
Using our attractor equations we may identify the constant $c$ as an element of the mass matrix  of the axino-dilatino $\phi$ and the complex structure of the torus, $\tau$:
\begin{equation}
  c= M_{\underline \tau \underline \phi} \ .
\label{mass}
\end{equation}
Here we can see the appearance of the new term in the  attractor equation with vanishing critical value of the central  charge, $e^{K\over 2}W=M_{3/2}=0$  which allows Minkowski vacuum with fixed moduli. 

Also in the case considered in \cite{Aspinwall:2005ad} the flux stabilizing all complex structure on K3 is a primitive (2,2) form so that the superpotential and the central charge vanish in the vacuum, only the second term in the attractor equation is valid and we find that in 
\begin{equation}
  F_4= 2 \rm Re (c \,  \bar \Omega^1 \wedge   \Omega^2)
\label{AK}
\end{equation}
the constant $c$ is again a mass of chiral fermions in the vacuum.


\subsection{Non-supersymmetric New  Flux Vacua Attractors}

The non-supersymmetric flux vacua attractors can be defined in complete analogy with all reasoning in the previous sections of the paper. Thus we expect that the extremal value of N=1 supergravity potential $V= |DZ|^2-3 |Z|^2 $ is possible at $DW\neq 0$. At the vacuum all moduli stop their running behavior.  We find the attractor equation  from the Hodge decomposition (\ref{G4})
\begin{equation}
  F_4= 2 \rm Re \left [\bar Z\hat \Omega_4 - \bar D^A \bar Z \hat D_A\Omega_4 + \bar D^{\underline 0 I}\bar Z \,  D_{\underline 0 I} \hat \Omega_4 \right ]_{DV=0}
\label{F4nonsusy1}
\end{equation}
The values of  moduli at the vacua have to be such   that the rhs of attractor equations is defined by the integer fluxes on the lhs.

One can expand eq. (\ref{F4nonsusy1}) in a way it was done for supersymmetric attractors and we find that
\begin{equation}
  F_4= 2 \rm Re \left [\bar Z\; \hat \Omega_1\wedge \hat \Omega_3 - \bar D^{\underline 0}  \bar Z \; \hat{\bar  \Omega} _1 \wedge \hat \Omega_3 - \bar D^{ I}  \bar Z \; \hat{  \Omega} _1 \wedge \hat \chi_I - \bar D^{\underline 0 I}\bar Z \;   \hat {\bar \Omega}_1\wedge \chi_I \right ]_{DV=0}
\label{F4nonsusy2}
\end{equation}
This equation can be made explicit using the equation $DV=0$ in the form $2  D_i Z\, \bar Z = D_i D_j Z \,  G^{j\bar j} \, \bar D_{\bar j} \bar Z$.

\section{Summary}

In this paper we derived new attractor equations defining supersymmetric flux vacua in IIB string theory. We have also  presented some well known equations for black hole attractors in the form in which they are easy to compare with the new attractors for flux vacua.   Finally, we proposed a new set of attractor equation for non-supersymmetric black holes and  flux vacua.

For supersymmetric black hole attractors the condensed form of attractor equations \cite{Ferrara:1995ih}-\cite{Ferrara:1996um} is given by 
 \be
h= 2 \rm Im \, (Z \; \bar { \Pi})_{DZ=0} \ ,
\label{BH}\ee
where the covariantly symplectic section  $\Pi$ is defined in eq. (\ref{Pi}). 
Using electric and magnetic charges $h= (p_h^\Lambda, q_{h\Lambda})$ we get
\begin{equation}
 \left (\matrix{
p_h^\Lambda\cr
q_{h\Lambda}\cr
}\right )= 2 {\rm Im}  \left(\matrix{
 Z \; \bar L^\Lambda\cr
  Z \; \bar M_\Lambda\cr
}\right )_{DZ=0} \ , \qquad Z=  L^\Lambda q_{h\Lambda}- p_h^\Lambda M_\Lambda \ . \label{stabBH}
\end{equation}
 Attractor equations define the minimum of the BPS black hole mass $M_{BH}^2= |Z|^2(t, \bar t, p, q)$. At the attractor point (\ref{stabBH}) where $t=t(p,q)$, $\bar t=\bar t(p,q)$ one finds that $|Z|^2_{DZ=0} = \pi S(p,q)$ where $S$ is the  entropy of the extremal BPS black hole $S={A\over 4}$ and $A$ is the area of the black hole horizon.

We found the following simple equation for non-supersymmetric black hole attractors:
\be
h= 2 {\rm Im} \left [  \;  Z \, \bar  \Pi -    D^{ I}Z \;
\bar D_{ I} {\bar \Pi} \right ]_{ \partial V_{BH} =0} \ , 
\label{nonsusy}\ee   
where $M_{BH}^2 = V_{BH}= |DZ|^2 + |Z|^2> |Z|^2$. The area of the horizon is proportional to the minimal value of this potential ${A\over 4\pi} =V_{BH}$  at $\partial V_{BH} =0$ \cite{Ferrara:1997tw,Goldstein:2005hq}. It would be interesting to check if the examples of non-supersymmetric black hole attractors recently studied in \cite{Goldstein:2005hq} satisfy  equation (\ref{nonsusy}). Some examples of non-supersymmetric black hole attractor equations have been already checked in \cite{AG} and confirm these equations.

We derived here the  new attractor equations describing the supersymmetric flux vacua in IIB string theory. They are given in the most condensed form by equation
\begin{equation}
  F_4= 2 \rm Re \left [\bar Z \; \hat \Omega_4 + \bar D^{\underline 0 I}\bar Z \;  D_{\underline 0 I} \hat \Omega_4 \right ]_{DZ=0} \ ,
\label{F4}
\end{equation}
with notations  explained near Eq. (\ref{G4}). In the next level of condensed notation, they are defined for the  doublet of fluxes in type IIB theory, $H_3$ and $F_3$ $SL(2, Z)$, $\tau$ being the axion-dilaton:
\begin{equation}
  \left (\matrix{
h\cr
\cr
f\cr
}\right )=   \left (\matrix{
 \, 2 \rm Re \, ( Z \; \bar \Pi) \cr
\cr
 \,   2 \rm Re (  Z \; \bar \tau \; \bar \Pi )\cr
}\right )_{DZ=0}+ 
  \left (\matrix{
 \, 2 \rm Re (\bar Z^{\underline 0I} \;  D_I \, \Pi)\cr
\cr
 \, 2 \rm Re ( \bar Z^{\underline 0I} \;\bar \tau \; D_I \Pi ) \cr
}\right )_{DZ=0}
\label{IIBshortconcl}
\end{equation}
Here $|Z|^2 = |M_{3/2}|^2$ is the square of the gravitino mass at the flux vacua and $D_{\underline 0 I} Z= M_{\underline 0 I}$ is the mass matrix of the axino-dilatino mixing with complex structure modulino in the vacua. Attractor equations (\ref{F4}),(\ref{IIBshortconcl}) define the supersymmetric extremum of the potential of N=1 supergravity, $V=e^K(|DW|^2-3 |W|^2)= |DZ|^2-|Z|^2$. These equations have been recently confirmed in \cite{AG} for a large class of supersymmetric flux vacua in type IIB string theory. 

We also found the attractor equation defining the  non-supersymmetric flux vacua in type IIB string theory. The most condensed form of this equation is 
\begin{equation}
  F_4= 2 \rm Re \left [\bar M_{3/2} \hat \Omega_4 - \bar F^A  \hat D_A\Omega_4 +\bar M^{\underline 0 I}\bar Z \,  D_{\underline 0 I} \hat \Omega_4 \right ]_{DV=0} \ ,
\label{F4nonsusy}
\end{equation}
where $F_A= D_A Z$. A more detailed form of this equation is given in (\ref{F4nonsusy2}), it adds two new terms to equation (\ref{IIBshortconcl}) proportional to the auxiliary F-terms in the axion-dilaton multiplet $F_{\underline 0}$ and complex structure moduli $F_I$.
Here again, it would be interesting to find if the non-supersymmetric flux vacua minimizing $V=e^K(|DW|^2-3 |W|^2)$  satisfy the attractor equation (\ref{F4nonsusy}), e. g. in examples of \cite{Saltman:2004sn}.

It is instructive to compare the supersymmetric black hole attractor equation (\ref{BH}) with supersymmetric flux vacua new attractors (\ref{IIBshortconcl}). The first term in (\ref{IIBshortconcl}) reminds (\ref{BH}), however Im is replaced by Re, the reason being that in the black hole  case we have a holomorphic 3-form on CY 3-fold, in the flux vacua case we have an effective holomorphic 4-form since the manifold has a CY 3-fold as well as an auxiliary torus which together make a 4-fold.

Even more significant difference is in the presence of the second term in eq. (\ref{IIBshortconcl}) which is totally absent in the black hole case. It allows stabilization of moduli even if  $|Z|_{DZ=0}=0$.
For regular supersymmetric black holes in classical N=2 supergravity,  the square of the central charge  $|Z|^2$ at the attractor point is proportional to the black hole entropy/area of the horizon, which at the attractor point depends only on charges and does not depend on continuous moduli. Typically the entropy is given by a product of charges. If one of the charges in this product vanishes, the area of the horizon vanishes, the metric has a null singularity, and some of the moduli may blow up. It was therefore important in the studies of supersymmetric black holes near the horizon to find cases with  central charge which does not vanish at the attractor point.\footnote{Extremal black holes with $Z=0$ require some special treatment, see e. g. \cite{Denef:2001xn}, \cite{Dabholkar:2004dq}, where the higher order corrections to the area and the entropy discovered in  \cite{LopesCardoso:1998wt} play a crucial role.} The function which was minimized was $|Z(t, \bar t, p, q)|^2$ with $t, \bar t$ taking arbitrary values in the moduli space corresponding to the values of these moduli at the asymptotic infinity. At the minimum, this function was proportional to the area of the horizon, $ {A(p, q)\over 4\pi}=  |Z(z(p,q), \bar z(p,q), p, q)|^2$.

In flux vacua we are looking for the critical points of the potential $V= |DZ|^2- 3 |Z|^2$. In this case, unlike in the case of extreme black holes, one may have solutions with all moduli stabilized  (axion-dilaton and complex structure) in flux vacua with $DZ=0$,  $Z=0$ and $V_{fix}=0$.  (The cases of flux vacua with $DZ=0$,  $Z=0$ and $V_{fix}=0$ have solutions with all moduli stabilized  (axion-dilaton and complex structure), as different from the situation with extremal black holes.) 
As we have shown in this paper,  such supersymmetric vacua can be described by the generalized  attractor equations. At the critical point $DZ=0$, the potential  is equal to $V=-3m_{3/2}^2=-3|Z|^2$, where $|Z|$ may or may not be  equal to zero.

Finally, it is amazing that simple  non-supersymmetric attractor equations, (\ref{nonsusy}), (\ref{F4nonsusy}), defining the black hole   and  flux vacua  are also available.  Despite the first discussion of non-supersymmetric black hole attractors was presented in \cite{Ferrara:1997tw}, not much attention was given to this case. Due to the recent intensive work on non-supersymmetric attractors in \cite{Goldstein:2005hq}, \cite{Tripathy:2005qp}  it become clear that, as always in dynamical systems, a simple universal description of the critical points may exist. This led us to the derivation of eqs.  (\ref{nonsusy}), (\ref{F4nonsusy}). We hope that more investigations will be performed in this direction, as it was done with supersymmetric attractors in the past.

\

\leftline{\bf Acknowledgments}

\

I am  grateful to P. Aspinwall, A. Ceresole,   G. Dall'Agata, R. D'Auria,  F. Denef, M. Douglas, S. Ferrara, B. Florea, 
S. Kachru, A. Linde, A. Maloney, G. Moore, M. Sasaki,  M. Trigiante, S. Trivedi, P. Yi and D. Waldram for
valuable discussions of flux vacua and attractors. I am particularly grateful  to   A. Giryavets for his work on examples of new attractors.  I am grateful for the hospitality to the organizers of the conference ``The Next Chapter in Einstein's Legacy'' at the Yukawa Institute, Kyoto,  and to the organizers and participants of the KIAS-YITP Joint workshop in Seoul ``Cosmological Landscape: Strings, Gravity, and Inflation'' where this work was   initiated.
It was supported by NSF grant 0244728 and by Kyoto University.

\end{document}